\documentclass[iop]{emulateapj}

\usepackage{hyperref}
\usepackage{xcolor}
\usepackage{url}
\usepackage{amssymb}
\usepackage{physics}
\hypersetup{
    colorlinks,
    linkcolor={red!50!black},
    citecolor={blue!50!black},
    urlcolor={black} 
}
\usepackage{mathrsfs}
\usepackage{nicefrac}

\newcommand{\feh}{\ensuremath{[\mbox{Fe}/\mbox{H}]}}

\shorttitle{Multi-Epoch Simulations of HD187123b}
\shortauthors{Buzard et al.}

\begin{document}

\title{Simulating the Multi-Epoch Direct Detection Technique to Isolate the Thermal Emission of the Non-transiting Hot Jupiter HD187123b}

\author{Cam Buzard\altaffilmark{1}, Luke Finnerty\altaffilmark{2}, Danielle Piskorz\altaffilmark{3}, Stefan Pelletier\altaffilmark{4}, Bj{\"o}rn Benneke\altaffilmark{4}, Chad F. Bender\altaffilmark{5}, Alexandra C. Lockwood\altaffilmark{6}, Nicole L. Wallack\altaffilmark{3}, Olivia H. Wilkins\altaffilmark{1}, Geoffrey A. Blake\altaffilmark{1,3}}

\altaffiltext{1}{Division of Chemistry and Chemical Engineering, California Institute of Technology, Pasadena, CA 91125}
\altaffiltext{2}{Division of Physics, Mathematics, and Astronomy, California Institute of Technology, Pasadena, CA 91125}
\altaffiltext{3}{Division of Geological and Planetary Sciences, California Institute of Technology, Pasadena, CA 91125}
\altaffiltext{4}{Institute for research on exoplanets, Universit{\'e} de Montr{\'e}al, Montreal, QC}
\altaffiltext{5}{Department of Astronomy and Steward Observatory, University of Arizona, Tucson, AZ 85721}
\altaffiltext{6}{Space Telescope Science Institute, Baltimore, MD 21218}

\begin{abstract}
We report the 6.5$\sigma$ detection of water from the hot Jupiter HD187123b with a Keplerian orbital velocity $K_p$ of 53 $\pm$ 13 km/s. This high confidence detection is made using a multi-epoch, high resolution, cross correlation technique, and corresponds to a planetary mass of 1.4$^{+0.5}_{-0.3}$ $M_J$ and an orbital inclination of 21 $\pm$ 5$^{\circ}$. The technique works by treating the planet/star system as a spectroscopic binary and obtaining high signal-to-noise, high resolution observations at multiple points across the planet's orbit to constrain the system's binary dynamical motion. All together, seven epochs of Keck/NIRSPEC $L$-band observations were obtained, with five before the instrument upgrade and two after. Using high resolution SCARLET planetary and PHOENIX stellar spectral models, along with a line-by-line telluric absorption model, we were able to drastically increase the confidence of the detection by running simulations that could reproduce, and thus remove, the non-random structured noise in the final likelihood space well. The ability to predict multi-epoch results will be extremely useful for furthering the technique. Here, we use these simulations to compare three different approaches to combining the cross correlations of high resolution spectra and find that the \citealt{zucker2003} log(L) approach is least affected by unwanted planet/star correlation for our HD187123 data set. Furthermore, we find that the same total S/N spread across an orbit in many, lower S/N epochs rather than fewer, higher S/N epochs could provide a more efficient detection. This work provides a necessary validation of multi-epoch simulations which can be used to guide future observations and will be key to studying the atmospheres of further separated, non-transiting exoplanets.
\end{abstract}


\section{Introduction}

To date, over four thousand extrasolar planets have been discovered with a range of vastly different orbital and atmospheric properties. The most detailed follow-up characterizations of these planets have been provided by the transit technique. While the transit technique can give invaluable insight into the atmospheres of these planets \citep[e.g.,][]{Madhusudhan2014}, it is restricted to systems with a very narrow range of orbital inclinations that allow them to transit with respect to our line-of-sight from Earth. While $\sim$10\% of typical hot Jupiters around Sun-like stars can be expected to transit, as we move to habitable zone planets around M stars and Sun-like stars, the transit probabilities drop to $\sim$9\% and 0.5\%, respectively. Direct imaging has also provided information on the atmospheric content and relative molecular abundances of planets at large separation \citep[e.g.,][]{Konopacky2013}, but these techniques are not yet sensitive to planets within $\sim$0.1" \citep[e.g.,][]{Snellen2014,Schwarz2016}, which excludes habitable zone planets around even the closest M stars. 

\begin{figure}
    \centering
    \noindent\includegraphics[width=21pc]{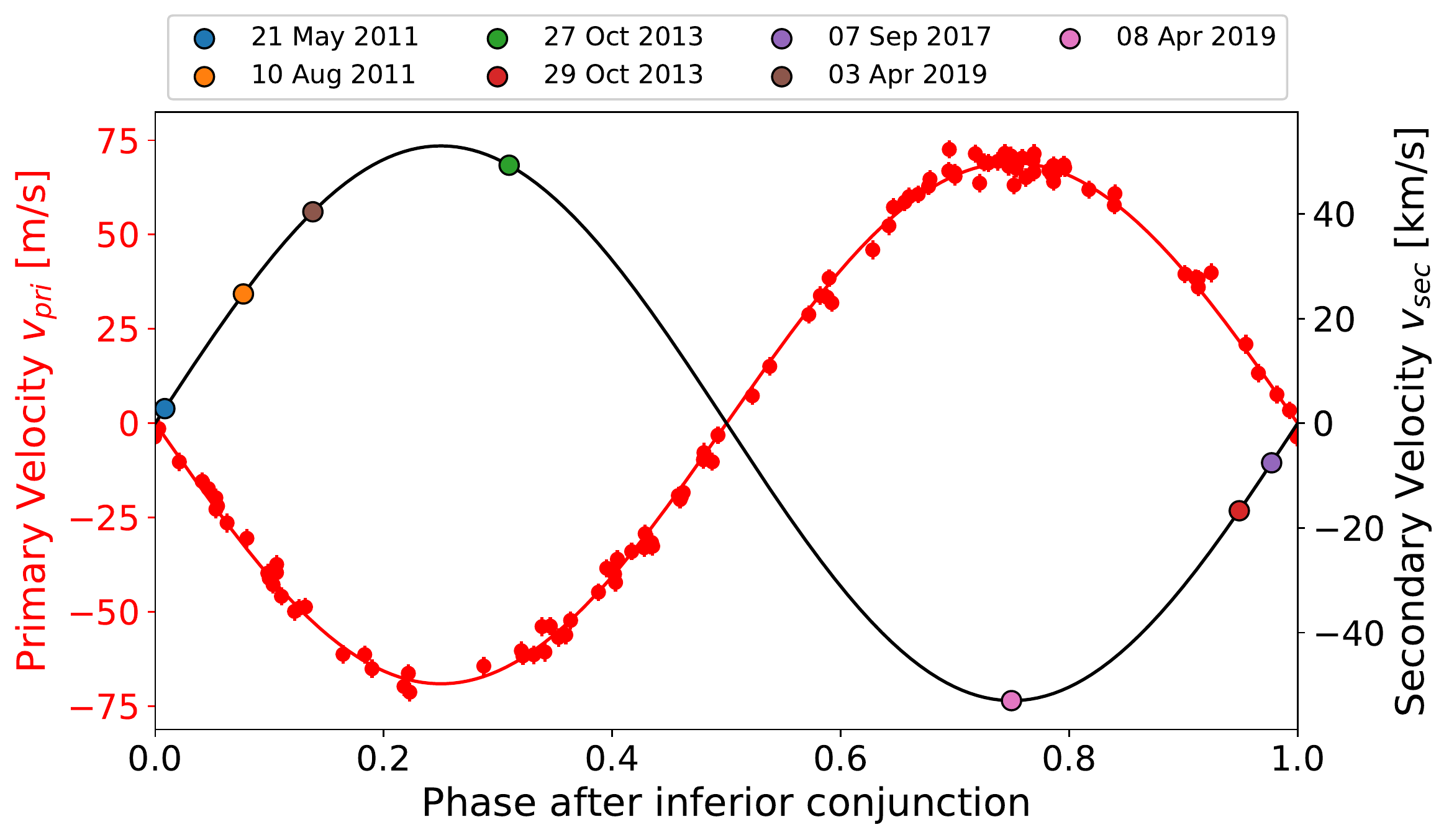}
    \caption{Model showing the spectroscopic binary nature of the HD187123 system. The red curve and points show the stellar radial velocity model and measurements \citep{Feng2015}, respectively, and the black curve shows the planetary velocity signature with the colored circles showing the planet's phase at each of our observations with $v_{sec}$ given by our measured $K_p$ of 53 km/s. }
    \label{RVcurve}
\end{figure}

Recent work has developed high resolution cross correlation techniques that aim to target the much larger sample of non-transiting, yet close-in, planets by separating the stellar and planetary signals by radial velocity rather than by flux variation, as in the transit technique, or by spatial separation, as in the direct imaging technique \citep[e.g.,][]{Snellen2010,lockwood}. These direct detection techniques work by treating a star/planet system as a spectroscopic binary and measuring the radial velocity signature of the planet. This signature will have an opposite phase to the stellar radial velocity curve (see Figure~\ref{RVcurve}), and by combining its amplitude, which we call $K_p$, the planetary Keplerian line-of-sight velocity, with the stellar radial velocity amplitude $K$, we can break the mass/inclination degeneracy left by the stellar radial velocity technique and further characterize the planet's atmosphere \citep[e.g.,][]{Brogi2012,Brogi2013,Brogi2014,lockwood,piskorz88133,piskorzupsand,Birkby2017,Piskorz2018}. These techniques have been used to detect the presence of H$_2$O \citep[e.g.,][]{Birkby2017}, CO \citep[e.g.,][]{Brogi2012}, TiO \citep{Nugroho2017}, HCN \citep[e.g.,][]{Hawker2018}, and CH$_4$ \citep{Guilluy2019} in planetary atmospheres, as well as winds \citep{Snellen2010} and planetary rotation rate \citep{Brogi2016}. They have been applied using data from VLT/CRIRES \citep[e.g.,][]{Snellen2010}, Keck/NIRSPEC \citep[e.g.,][]{lockwood}, ESO/HARPS \citep[e.g.,][]{Martins2015}, CFHT/ESPaDOnS \citep[e.g.,][]{Esteves2017}, GIANO \citep[e.g.,][]{Brogi2018}, and CARMENES \citep[e.g.,][]{AlonsoFloriano2019} to study about 10 hot Jupiters.

There are two main methods that have been applied to measure planetary Keplerian orbital velocites $K_p$: a single-night version and a multi-epoch version. The single-night version \citep[e.g.,][]{Snellen2010} observes the systems over a full night ($\sim$5-7 hours) when the planet is near superior or inferior conjunction, where its line-of-sight velocity changes most rapidly, and watches for the planetary lines to move across detector pixels as the stellar and telluric lines remain stationary. This technique can also be applied to multiple partial nights as long as the planet lines move across the detector's pixels in the partial nights \citep[e.g., HD 179949, ][]{Brogi2014}. The single-night method has provided several high confidence detections of planetary emission and molecular features, but requires the planetary lines to move by tens of km/s over a $\sim$5-7 hour night, and so is limited to close-in planets. The multi-epoch method \citep[e.g.,][]{lockwood}, rather than looking for shifting planetary lines in a single night, observes at multiple epochs around the planet's orbit for $\sim$2-3 hours per epoch. These times are chosen to be long enough to maximize the signal-to-noise on the system and to allow for a principal component analysis telluric correction (as described in Section~\ref{reduction}) but short enough that the planetary lines stay fixed, and so are not removed by the telluric correction. Because the multi-epoch technique does not require the planetary lines to move in a short time period, it is applicable to the future study of planets at larger orbital radii, including those in habitable zones. It could study planets in M dwarf habitable zones out to those in K dwarf and solar habitable zones that are too far out for the single-night method but too close in for direct imaging techniques with current adaptive optics capabilities.   
 
As such, improvements on the multi-epoch technique are timely and critical. Here, we apply the multi-epoch method to the hot Jupiter HD187123b, using simulations to understand the limiting factors in our detection. As one of only two known systems with a hot Jupiter (gas giant with $P < 10$ days and $M\sin i > 0.1 M_{Jup}$) and a very-long period planet ($P>5$ yrs) in a well determined orbit \citep{Feng2015}, this system could hold valuable clues to understanding planetary migration. The system is orbiting the Sun-like G2V star HD187123A. HD187123b, the hot Jupiter, has a minimum mass of 0.51 $M_{Jup}$ and an orbital period of 3.10 days. HD187123c is the Jupiter-analogue in the system. It is on an eccentric ($e$ = 0.280) orbit with a period of 9.1 yrs and a minimum mass of 1.8 $M_{Jup}$ \citep{Feng2015}. HD187123b was first discovered by \citealt{butler1998} and the most up-to-date Keck/HIRES radial velocity data set was analyzed by \citealt{Feng2015} (see Figure~\ref{RVcurve}). The relevant properties of HD187134A and HD187123b are given in Table~\ref{systemproperties}. 

In Section~\ref{obsreduc}, we describe the Keck/NIRSPEC data sets and their reduction. In Section~\ref{simulationsection}, we describe how we simulate multi-epoch data. We use our simulation framework to measure the $K_p$ of HD187123b along with its mass and inclination in Section~\ref{analysis}. We consider the trade-off between signal-to-noise (S/N) per epoch and orbital coverage in Section~\ref{snsection}, and discuss and conclude in Sections~\ref{discuss} and \ref{conclude}, respectively. 


\begin{deluxetable}{llc}
\tablewidth{0pt}
\tabletypesize{\scriptsize}
\tablecaption{HD187123 System Properties}
\tablehead{Property & Value & Ref.} 
\startdata
\sidehead{\textbf{HD187123A}}
Mass, $M_{\star}$ & 1.037 $\pm$ 0.025 $M_{\sun}$ & (1)  \\
Radius, $R_{\star}$ & 1.143 $\pm$ $0.039R_{\sun}$ & (2) \\
Effective temperature, $T_{\mathrm{eff}}$ & 5815 $\pm$ 44 K & (3) \\
Metallicity, \feh &0.121 $\pm$ 0.30 & (3) \\
Surface gravity, $\log g$ & 4.359 $\pm 0.060$ & (3) \\
Rotational velocity, $v \sin i$ & 2.15 $\pm$ 0.50 km/s & (3) \\
Systemic velocity, $v_{sys}$ & -16.965 $\pm$ 0.0503 km/s & (4) \\
\textit{K} band magnitude, $K_{mag}$ & 6.337 & (5) \\
\sidehead{\textbf{HD187123b}}
Velocity semi-amplitude, $K$ & 69.04 $^{+0.42}_{-0.43}$ m/s & (6) \\
Line-of-sight orbital velocity, $K_P$ & 53 $\pm$ 13 km/s & (6) \\
Minimum mass, $M_p\sin i$ & 0.5077 $^{+0.0087}_{-0.0088}$ $M_J$ & (6) \\
Mass, $M_p$ & 1.4$^{+0.5}_{-0.3}$ $M_J$ & (6) \\
Inclination, $i$ & $21 \pm 5^{\circ}$ & (6) \\
Semi-major axis, $a$ & 0.04209 $\pm$ 0.00034 AU & (6) \\
Period, $P$ &3.0965885 $^{+0.0000051}_{-0.0000052}$ days & (6) \\
Eccentricity, $e$ & 0.0076 $^{+0.0060}_{-0.0049}$ & (6) \\
Time of periastron, $T_{peri}$ &   2454342.87 $\pm 0.30$ JD &(2) \\
Argument of periastron, $\omega$ &  360 $\pm$ 200$^{\circ}$   & (2) \\
Time of inferior conjunction, $T_{o}$ & 2454343.6765$^{+0.0064}_{-0.0074}$ JD &(6) 
\enddata
\label{systemproperties}
\tablerefs{(1) \citealt{Takeda2007}, (2) \citealt{Feng2015}, (3) \citealt{valenti2005}, (4) \citealt{Soubiran2013}, (5) \citealt{Cutri2003}, (6) This work} 
\end{deluxetable}


\section{NIRSPEC Observations and Data Reduction}
\label{obsreduc}
\subsection{Observations}
\label{nirspecobs}
We observed the HD187123 system for seven nights in the $L$ band using NIRSPEC (Near InfraRed SPECtrometer; \citealt{McLean1998}) at the Keck Observatory. Two of the nights were obtained with the upgraded NIRSPEC instrument \citep{martin2018}, while the rest were taken with the original. We used an ABBA nodding pattern and obtained spectral resolutions of $\sim$25,000 pre-upgrade with the 0.432"$\times$24" slit setup and $\sim$41,000 in $L$ post-upgrade with the 0.288"$\times$24" slit setup. Before the instrument upgrade, we used echelle settings to obtain orders typically covering 3.4022-3.4550, 3.2549-3.3055, 3.1200-3.1685, 2.9959-3.0424 $\mu$m. Our post-upgrade $L$ band settings covered 3.6292-3.6965, 3.4630-3.5292, 3.3131-3.3764, 3.1758-3.2364, 3.0495-3.1075, 2.9330-2.9886 $\mu$m. Note that the band settings before and after the upgrade do not overlap. Table~\ref{observationtable} gives the details of these observations.

\begin{deluxetable*}{lccccc} 
\tablewidth{0pt}
\def\arraystretch{1}
\tablecaption{NIRSPEC Observations of HD187123}
\tablehead{Date & Julian Date$^{\tablenotemark{a}}$ & Shifted mean  & Barycentric velocity  & Integration time & S/N$^{\tablenotemark{c}}_{\textit{L}}$ \\
 & ($-2,400,000$ days) &  anomaly $M'^{\tablenotemark{a,b}}$  & $v_{bary}$ (km/s) & (min) & }
\startdata
2011 May 21  & 55703.105 & 0.01 & 16.16 & 56 & 1724\\ 
2011 Aug 10  & 55783.829 & 0.08 & -2.48 & 108 & 1713\\ 
2013 Oct 27  & 56592.759 & 0.31 & -17.44 & 44 & 1283\\ 
2013 Oct 29  & 56594.738 & 0.95 & -17.50 & 80 & 2050\\ 
2017 Sep 7  & 58003.774 & 0.98 & -10.15 & 96 &2409 \\
2019 Apr 3$^{\tablenotemark{d}}$  & 58577.140 & 0.14 & 15.49 & 84 & 2298\\
2019 Apr 8$^{\tablenotemark{d}}$  & 58582.131 & 0.75 & 16.09 & 64 & 3417
\enddata
\label{observationtable}
\tablenotetext{a}{Julian date and shifted mean anomaly refer to the middle of the observing sequence.}
\tablenotetext{b}{We report a shifted mean anomaly ($M'$) that is defined from inferior conjunction, rather than from the pericenter, and runs from 0 to 1. }
\tablenotetext{c}{S/N$_{\textit{L}}$ is calculated at 3.0 $\mu$m. Each S/N calculation is for a single channel (i.e., resolution element) for the whole observation.}
\tablenotetext{d}{These observations were taken with the upgraded NIRSPEC instrument.}
\end{deluxetable*}

\subsection{NIRSPEC Data Reduction}
\label{reduction}
We reduce our NIRSPEC data using the Python pipeline described by \citealt{piskorz88133}, adapting the pipeline where necessary to reduce the 2 nights of data from the upgraded NIRSPEC instrument. The two-dimensional images are flat-fielded and dark subtracted according to \citealt{Boogert2002}. The extracted one-dimensional spectra are then wavelength calibrated with a fourth-order polynomial fit according to model telluric lines. 

\begin{figure}
    \centering
    \noindent\includegraphics[width=21pc]{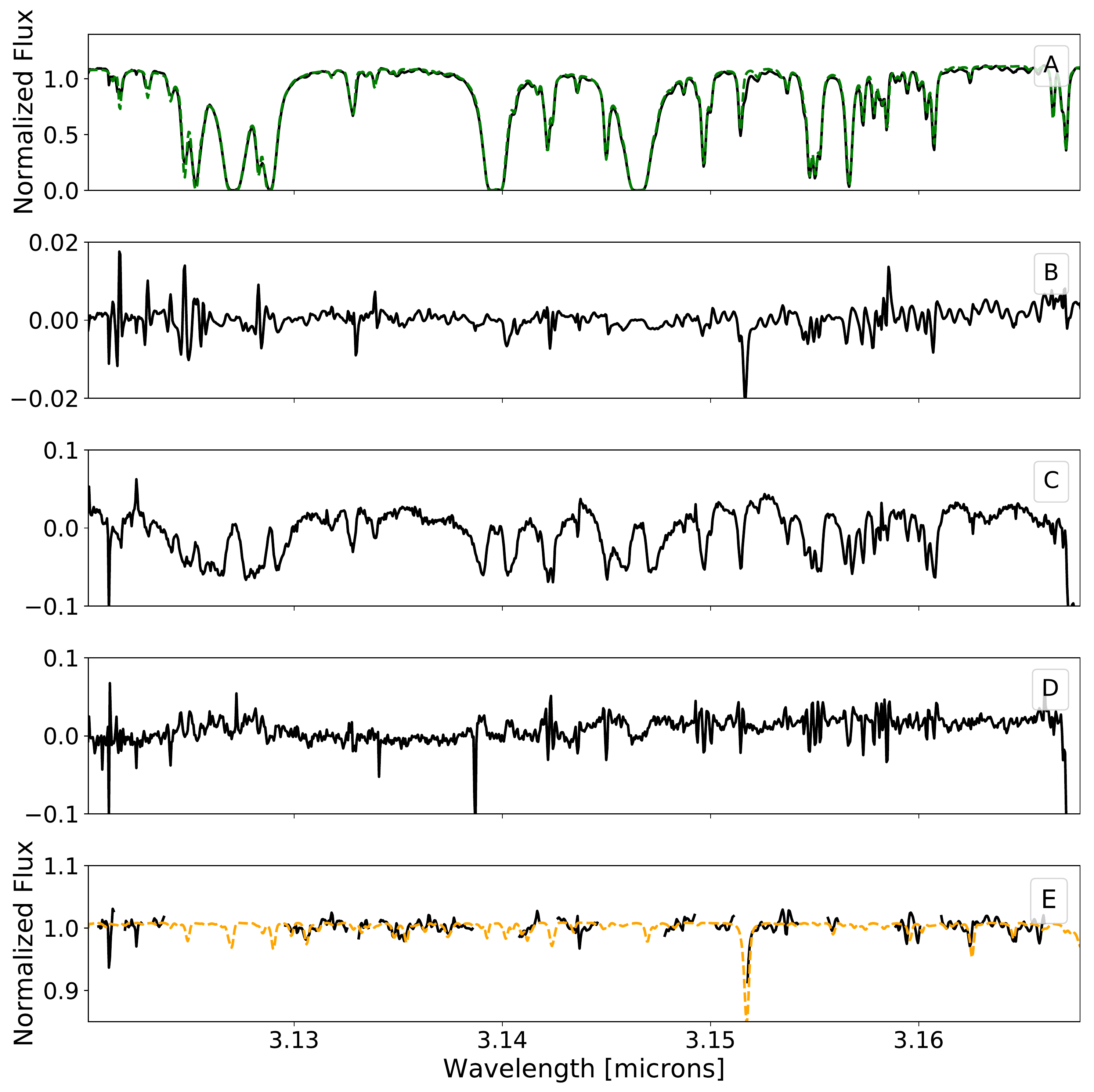}
    \caption{Demonstration of PCA telluric removal approach. (A): Raw spectrum of HD187123 from September 7, 2017 with the initial telluric model fit shown in green. (B-D): The first three principal components identified in arbitrary units. These describe changes in the airmass, molecular abundances in the Earth's atmosphere, and plate scale over the course of the observation. (E): Same as A, but without the initial telluric model fit and the first five principal components. A stellar model is overplotted in orange. }
    \label{pcafigure}
\end{figure}

After the 1-D spectra are extracted and wavelength-calibrated, a model-guided principal component analysis (PCA) is used to remove time-variable components from the data. We use the ESO tool \texttt{Molecfit} \citep{Kausch2014} to fit the initial telluric model to each night of data. In addition to fitting the telluric abundances and continuum, \texttt{Molecfit} uses a Gaussian fit to determine the resolution of the data. It reports the full-width at half maximum (FWHM) of the Gaussian kernel, which we later use to broaden the stellar and planetary templates for cross correlation. After the best fit model is removed from each nod in the data set, PCA is used to identify the dominant sources of variance, following the technique developed in \citealt{piskorz88133}. Typically, the majority of the variance is accounted for in the first few principal components. These components typically contain variance due to changes in telluric abundances, in airmass, in the continuum, and in instrument response. After these first few components are removed, a clean stellar/planetary spectrum is left behind. Figure~\ref{pcafigure} shows the third order of the data from Sep 7, 2017 with its initial telluric fit, the first three principal components, and the clean stellar+planetary spectrum. We specifically limit our observation times so that the planetary signal does not move across pixels in the course of a single night observation, to ensure that PCA will not remove the planetary signal. For the rest of this work, we use spectra with three components and five fringes removed. We also mask out pixels in which telluric absorption features are stronger than 25\%. This results in between 9 and 68\% of each order being lost. Panel E of Figure~\ref{pcafigure} shows an order from Sep. 7, 2017 with these regions masked out.

\section{Simulating NIRSPEC Observations} \label{simulationsection}
After telluric correction, we use a two-dimensional cross correlation technique to detect the planetary velocity each night. Because of the difficulty in detecting the planetary velocity in only one epoch, due to the planet's low contrast relative to the star, the correlations from the different nights are combined. This is what allows us to detect the true planetary line-of-sight Keplerian orbital velocity. In order to run the cross correlation, we need high resolution, high fidelity stellar and planetary spectral models. We also need a reliable method of combining the correlations from different nights. Before describing the analysis of our HD187123b data, we first describe the spectral models used for the cross correlation in Section~\ref{models} and describe how we simulate the data at each epoch to help determine the true planetary velocity in Section~\ref{describingsimulations}. We describe the math behind the three different approaches to combining cross correlations in the Appendix.


\subsection{High Resolution Spectral Models}
\label{models}

We use an R = 250,000 high-resolution thermal emission model of HD187123b generated using the SCARLET framework \citep{benneke}. The model computes both the equilibrium chemistry and temperature structure of HD187123b assuming a solar elemental composition, perfect heat redistribution, and an internal heat flux of 75 K. The spectrum is calculated assuming an atmosphere with a metallicity equal to that of the Sun and a C/O ratio of 0.54. The default temperature structure used in this work is inverted due to the inclusion of short wavelength absorbers TiO and VO. The SCARLET model framework includes molecular opacities of H$_2$O, CH$_4$, HCN, CO, CO$_2$, NH$_3$ and TiO from the ExoMol database (\citealt{Tennyson2012}), molecular opacities of O$_2$, O$_3$, OH, C$_2$H$_2$, C$_2$H$_4$, C$_2$H$_6$, H$_2$O$_2$, and HO$_2$ (HITRAN database by \citealt{Rothman2009}), alkali metal absorptions (VALD database by \citealt{Piskunov1995}), H$_2$ broadening \citep{Burrows2003}, and collision-induced broadening from H$_2$/H$_2$ and H$_2$/He collisions \citep{Borysow2002}. We broaden the planetary model with the instrument profiles fit to the data. The $L$ band portion of the spectral model, covering our data, is dominated by water emission features. 

We use a stellar model obtained by interpolating PHOENIX models \citep{Husser2013} to the effective temperature $T_{\mathrm{eff}}$, surface gravity $\log(g)$, and metallicity [Fe/H] values for HD187123A listed in Table~\ref{systemproperties}. Instrumental broadening is ultimately determined by the size of the intrument's pixels. The original $L$ band NIRSPEC pixels covered $\sim$5 km/s, and the upgraded $L$ band pixels cover $\sim$3.1 km/s. Because HD187123A is a slow rotator, with a rotational velocity of only 2.15 km/s, instrumental broadening will dominate over rotational broadening and, as such, we broaden the stellar model with only the kernels determined in Section~\ref{reduction}.

\subsection{Simulating Multi-Epoch Data} \label{describingsimulations}
In this work, we simulate the multi-epoch data to better understand the strengths and weaknesses of the technique. To do this, we start with the high resolution SCARLET planetary and PHOENIX stellar models described in Section~\ref{models}. We scale each model by its surface area, i.e. multiply it by its radius squared. The stellar radius is well measured (see Table~\ref{systemproperties}), but because it is a non-transiting system, the planetary radius is not. We assume a radius of 1.0 R$_J$. With this planetary radius, the simulated data has an average spectroscopic planet/star contrast of $1.2\times10^{-3}$ in the $L$ band.


After the stellar and planetary models are appropriately scaled, they are shifted to the nightly velocities. The stellar spectrum is shifted by 
\begin{equation} \label{vpri}
    v_{pri} = v_{sys} - v_{bary}
\end{equation}
where $v_{sys}$ is the systemic radial velocity and $v_{bary}$ is the nightly barycentric velocity in the direction of the system. The planetary spectrum is shifted by 
\begin{equation} \label{vsec}
    v_{sec} = K_p\sin \bigg{(}\frac{2\pi}{P}(T_{obs}-T_o)\bigg{)} + v_{pri}
\end{equation}
where $K_p$ is the line-of-sight Keplerian velocity of the planet, $P$ is the orbital period, $T_o$ is the time of inferior conjunction, and $T_{obs}$ is the midpoint of the observation in Julian date. Unless otherwise stated, $P$, $T_o$, and $v_{sys}$ are set as the values in Table~\ref{systemproperties}. The $P$ and $T_o$ values reported were measured using \texttt{RadVel} \citep{radvel} to refit the radial velocity data from \citealt{Feng2015}. We measure equivalent values of $P$, $e$, $K$, and using the same stellar mass estimate from \citealt{Takeda2007}, $M_p\sin i$ and $a$ to those found in \citealt{Feng2015}. However, by refitting the data, we can directly measure the time of inferior conjunction, $T_o$, and its uncertainty. The uncertainty we measure on $T_o$ is only $\sim$0.2\% of the orbital period, meaning that we have a very good sense of where the planet is on its orbit during each epoch. While this would not make much of a difference to the detection ability of the simulations, it will be important for detecting the planet in the real data (described in Section~\ref{orbitsoln}). The $T_{obs}$ and $v_{bary}$ values are from Table~\ref{observationtable}. $K_p$ is a free parameter.

Next, the stellar model is linearly interpolated onto the planetary model wavelength axis and the two models are added. The stellar continuum is then removed using a $3^{\mathrm{rd}}$ order polynomial fit to the combined spectrum in wavenumber space from 2.8 to 4 $\mu$m. The stellar spectral template used to cross correlate the data (and simulated data) is continuum normalized in the same way (Section~\ref{corr}).

The spectra are then broadened according to the instrument profiles fit to the data and interpolated onto the wavelength axes for each of the orders and nights. The same pixels that are clipped from the data (described in Section~\ref{reduction}) are clipped from these simulated data as well. Lastly, random Gaussian noise is added to the simulated data at the level measured from the real data and reported in Table~\ref{observationtable}. 

These simulations account for sections of the data that have to be clipped, but assume that the PCA routine effectively removes all residual telluric structure from the data.

\section{NIRSPEC Data Analysis and Results}
\label{analysis}

We use two-dimensional cross correlations to determine the stellar and planetary velocities in each epoch of data. While the stellar velocities are readily apparent from single epochs, we must combine cross correlations from multiple epochs to detect the planetary velocity. Cross correlations can be combined as log likelihoods. Throughout this paper, we will call the process of converting cross correlations to log likelihoods ``CC-to-log(L)". \citealt{zucker2003} presented an approach to converting cross correlations into log likelihoods that can be applied in two ways which we will call the Zucker log(L) and Zucker maximimum likelihood (ML) approaches. \citealt{brogi2019} recently presented a new CC-to-log(L) approach. The math of these three approaches is described in the Appendix. We use each of these three approaches to combine the seven epochs of HD187123 data and compare the results each gives.

Now that we have presented the stellar and planetary spectral models and introduced the different CC-to-log(L) approaches, we describe our analysis of the HD187123b data. 

\subsection{Two-Dimensional Cross Correlation}
\label{corr}
\begin{figure}[t]
    \centering
    \noindent\includegraphics[width=21pc]{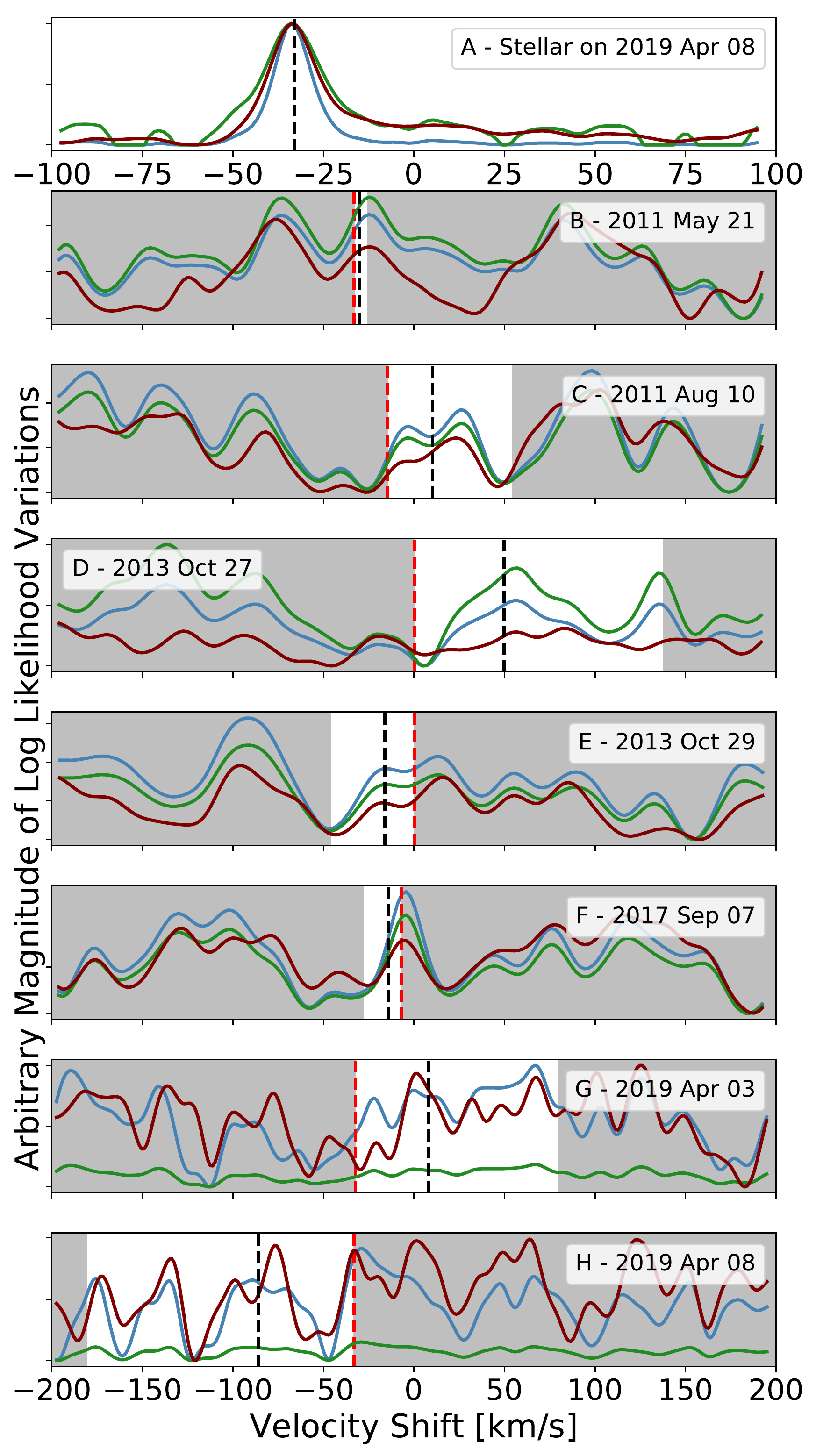}
    \caption{Log likelihood functions for all 7 epochs of NIRSPEC data on HD187123. (A): The stellar correlation from Apr 8, 2019. (B-H): The planetary likelihoods for each of the epochs. The colors represent different CC-to-log(L) approaches with Zucker log(L) in blue, Zucker ML in green, and Brogi \& Line in maroon. The curves are normalized, so the y-magnitude is arbitrary, but the relative heights between epochs combined the same way are maintained. The white regions show the allowable velocities, defined between face-on (red dashed line) and edge-on configurations, for each epoch given the known orbital position. The planetary mass/inclination of the system would determine where the planet would fall within the allowed regions.   }
    \label{mlcombine}
\end{figure}

We measure the stellar and planetary velocities using the two dimensional cross correlation technique (TODCOR, Equation~\ref{2dcc}) from \citealt{zuckertodcor} and the stellar and planetary spectral models described in Section~\ref{models}. In each night of data, we detect the star's velocity as expected (see Panel A of Figure~\ref{mlcombine}). Panels B-H of Figure~\ref{mlcombine} show the log likelihoods from each of the nights combined using each of the three CC-to-log(L) approaches: Zucker log(L) (blue), Zucker ML (green), and Brogi \& Line (maroon). The log likelihoods are normalized so that they fit on the same scale, but the relative heights of the log likelihoods between the nights for each CC-to-log(L) approach are maintained. The Zucker log(L) and Zucker ML log likelihoods have the same functional shapes, but the different nights are weighted differently. In each panel, the vertical dashed red line represents the velocity of the star during that epoch, which would correspond to the planetary velocity if the system were face-on. The white region, which illustrates the range of possible planetary velocities each night, begins there and extends until it reaches the maximum orbital velocity (given by $2\pi a/P$), which would represent an edge-on system. The planet's mass and inclination will determine where the peak will be within the white region. 

Panels G and H are from the NIRSPEC2 data. The increased resolution of the upgraded instrument can easily be seen in the more resolved structure in these panels as compared to Panels B-F. 


The sizes of the white regions also illustrate that some epochs have better constraining power than others. When the planet is near inferior or superior conjunction (M$\sim$0, 0.5), as on May 21, 2011, the nightly planetary velocity ($v_{sec}$) will be largely independent of $K_p$. When the planet is near quadrature (M$\sim$0.25, 0.75), however, as on Apr 8, 2019, the nightly planetary velocity changes significantly as a function of $K_p$. Thus, quadrature epochs are more useful for constraining $K_p$ than are those near conjuncture. We note that the opposite is true for the single-night technique. While the multi-epoch technique is most sensitive to epochs with the largest separation between the planetary and stellar velocities (i.e. quadrature), the single-night technique is most sensitive to orbital positions that give access to the largest change in planetary velocity over a short time period (i.e. near superior/inferior conjunction).

\subsection{Planet Mass and Orbital Solution}
\label{orbitsoln}

Because the planetary velocities cannot be reliably measured from single epochs, we combine the seven epochs to measure the $K_p$ of HD187123. As described in the Appendix, the log likelihoods from different epochs are combined by converting them from $v_{sec}$ to $K_p$ space using Equation~\ref{vsec} and then summing them.


\begin{figure}
    \centering
    \noindent\includegraphics[width=21pc]{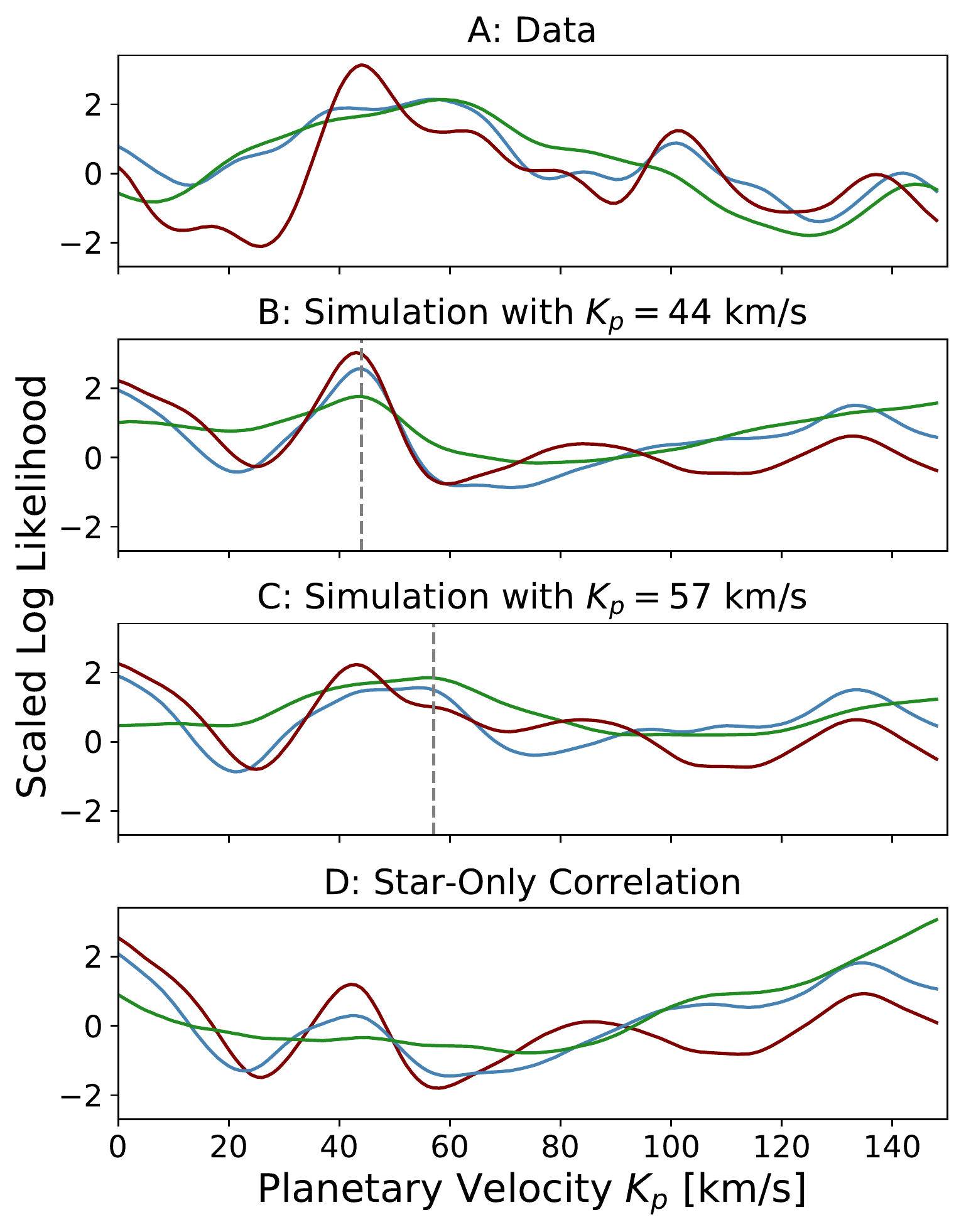}
    \caption{Normalized log likelihoods as a function of Keplerian orbital velocity $K_p$. The Zucker log(L), Zucker ML, and Brogi \& Line CC-to-log(L) combination techniques are shown in blue, green, and maroon, respectively. (A): The results of the data. (B-C): The results of the system simulations with a $K_p$ of 44 km/s and 57 km/s, respectively. These simulations both consider a 1 $R_{Jup}$ planet. The results of the simulations in Panel C match the data results in Panel A much better than do the simulation results in Panel B. (D): Similar to Figure~\ref{maxlike}B and \ref{maxlike}C, but with no injected planetary signal. All structure represents unwanted correlation between stellar model and planetary model spectral lines. }   
    \label{maxlike}
\end{figure}

Panel A of Figure~\ref{maxlike} shows the combined log likelihoods using the three different CC-to-log(L) methods. The three methods each produce a significant peak between around 45 to 60 km/s. To determine the correct Keplerian velocity, we simulate the effect of a 1.0 $R_J$ HD187123b-like planet at 44 and 57 km/s (shown in Panels B and C of Figure~\ref{maxlike}). We see that while both CC-to-log(L) approaches can uniquely detect the planet at 44 km/s, when the planet is shifted to 57 km/s, a side peak appears around 44 km/s. In the Brogi \& Line approach, this side peak is stronger than the real peak at 57 km/s while in both Zucker approaches the 57 km/s peak is broadened. We see a similar pattern when we compare these results to the log likelihoods derived from the data (Panel A). The Zucker log(L) approach shows two approximately equal height peaks at $\sim$40 and $\sim$57 km/s while the Brogi \& Line approach has a dominant peak at 44 km/s with a much weaker side peak at $\sim$63 km/s. 


Both sets of simulations also show a bump at around $\sim$135 km/s, which is also seen in the data. The \citealt{zucker2003} log(L) and \citealt{brogi2019} log(L) approaches do give rise to a small peak at about 100 km/s in the data that does not appear in the simulations. This side peak does not appear in the \citealt{zucker2003} ML approach on the data however. We therefore can rule out the peak at $\sim$100 km/s as the true planetary velocity. 

One difference between the simulated results and the data results is the magnitude of the log likelihood variation. We show scaled log likelihood curves in Figure~\ref{maxlike} so that the curves can be plotted on the same axes. In general, the variation in the simulated log likelihoods from -150 to 150 km/s is $\sim 5 \times$ the variation in the data log likelihood curves. We have found that varying the spectroscopic contrast $\alpha$, which is a function of the planetary radius, used to run the 2D cross correlation (described in the Appendix), changes the magnitude, but not the shape, of the resulting log likelihood curves. Therefore, the magnitude difference is likely due to the uncertainty in the planetary radius and lapse rate. We also note that the simulations seem to show a larger rise toward 0 km/s than is seen in the data. This is likely from correlation between the stellar component of the simulated data and the star model template that leaked into the second dimension of the correlation. In the simulated data, we use the same stellar model spectrum to generate the simulated data and to correlate it. In the real data on the other hand, the real stellar spectrum could be slightly different from the stellar spectral model used to correlate it. For instance, the stellar spectral model does not consider any starspots that could introduce a lower temperature component to the real stellar spectrum. The better match between the stellar template and the stellar component in the simulated data than in the real data would explain why the peak at 0 km/s is stronger in the simulated cases than in the real case.

There are several factors in addition to a lack of modeled starspots that could be leading to a discrepancy between our data results and our modeled results. One stems from inaccuracies in the molecular opacities in both our planetary and stellar spectral models. The ExoMol database uses the MARVEL \citep{Furtenbacher2007} procedure to correct theoretical calculations of transition frequencies and line shapes using laboratory experiments. The MARVEL framework has only been applied to a few molecules, however, including H$_2$O and TiO, but notably missing CH$_4$ and CO$_2$\footnote{\href{http://kkrk.chem.elte.hu/Marvelonline/molecules.php}{http://kkrk.chem.elte.hu/Marvelonline/molecules.php}}. The molecules not corrected by MARVEL have errors in transition frequencies around 0.1 cm$^{-1}$, which is around the resolution element of NIRSPEC. These errors, which are accounted for in the simulated results since the same planetary spectral model is used to generate the simulated data as to correlate it, are not accounted for in the real data and so could cause discrepancies between the two results. Inaccuracies in the stellar line lists could produce similar discrepancies. 

An additional source of discrepancy between the simulated and real results could be from our use of the literature value of $v_{sys}$ to combine the data from different epochs. Again, the same systemic (and barycentric) velocities are used to simulate the data as to cross correlation it. However, there are several sources \citep[e.g. rotation, winds,][]{Zhang2017} that are known to shift the real planetary emission a few km/s from the systemic velocity measured from star. We choose to only consider the planetary cut along the known stellar velocity, though, and so this could account for some discrepancy between the data and simulated results. 

We consider the peak at $\sim$57 km/s to be the true planetary detection. To test if we could determine where the extra correlation peaks, notably the one at 44 km/s come from, we ran additional simulations with no planet present in the simulated data. These are set up the same way as the simulations shown in the Panels B and C of Figure~\ref{maxlike}, but this time there is no planet model added in to the simulated data. We then run the two dimensional cross correlation, as above, and show the results of the combined planetary log likelihoods in Figure~\ref{maxlike}D. Because there is no planetary signal in the simulated data, the second dimension of the cross correlation, which involves correlating the data with a planetary model, shows the correlation between the stellar lines in the data and the planetary model. Figure~\ref{maxlike}D shows that this unintended star/planet correlation gives rise to both the peak at 44 km/s and the bump at $\sim$135 km/s. We also see from the flatness of the green curve that the Zucker ML approach is least affected by planet/star correlation. These results support our conclusion that the true $K_p$ is at 57 km/s rather than at 44 km/s. 


In general, we find that the two Zucker methods do not have as large peaks at incorrect values of $K_p$ as the Brogi \& Line method does for this data set. Figure~\ref{mlcombine} shows the log likelihoods computed for each epoch from each of the three combination approaches. We note that the Brogi \& Line method gives more weighting to the two NIRSPEC2 epochs (G, H) than to the five NIRSPEC1 epochs (B-F) while the Zucker log(L) approach gives more even weighting to all of the seven epochs. The Brogi \& Line combinations of the two NIRSPEC2 epochs each show a peak that corresponds to a $K_p$ of 44 km/s (just next to the black dashed lines in the direction of the red dashed lines in Panels G and H of Figure~\ref{mlcombine}), that does not appear in the five NIRSPEC1 epochs. Since the NIRSPEC1 and NIRSPEC2 $L$ band settings cover slightly different wavelength regions (see Section~\ref{nirspecobs}), this extraneous peak could be the result of correlation between stellar and planetary lines present in the NIRSPEC2 wavelength regions that are not in the NIRSPEC1 regions. Because the Brogi \& Line approach gives more weight to these epochs, the extraneous peak is not diluted by the NIRSPEC1 epochs as much as it is in the Zucker log(L) combination approach. On the other hand, the Zucker ML approach gives more weight to the NIRSPEC1 epochs than the NIRSPEC2 epochs, so does not benefit from the improved resolution of the NIRSPEC2 data in the same way that the Zucker log(L) results do.

\begin{figure}
    \centering
    \noindent\includegraphics[width=21pc]{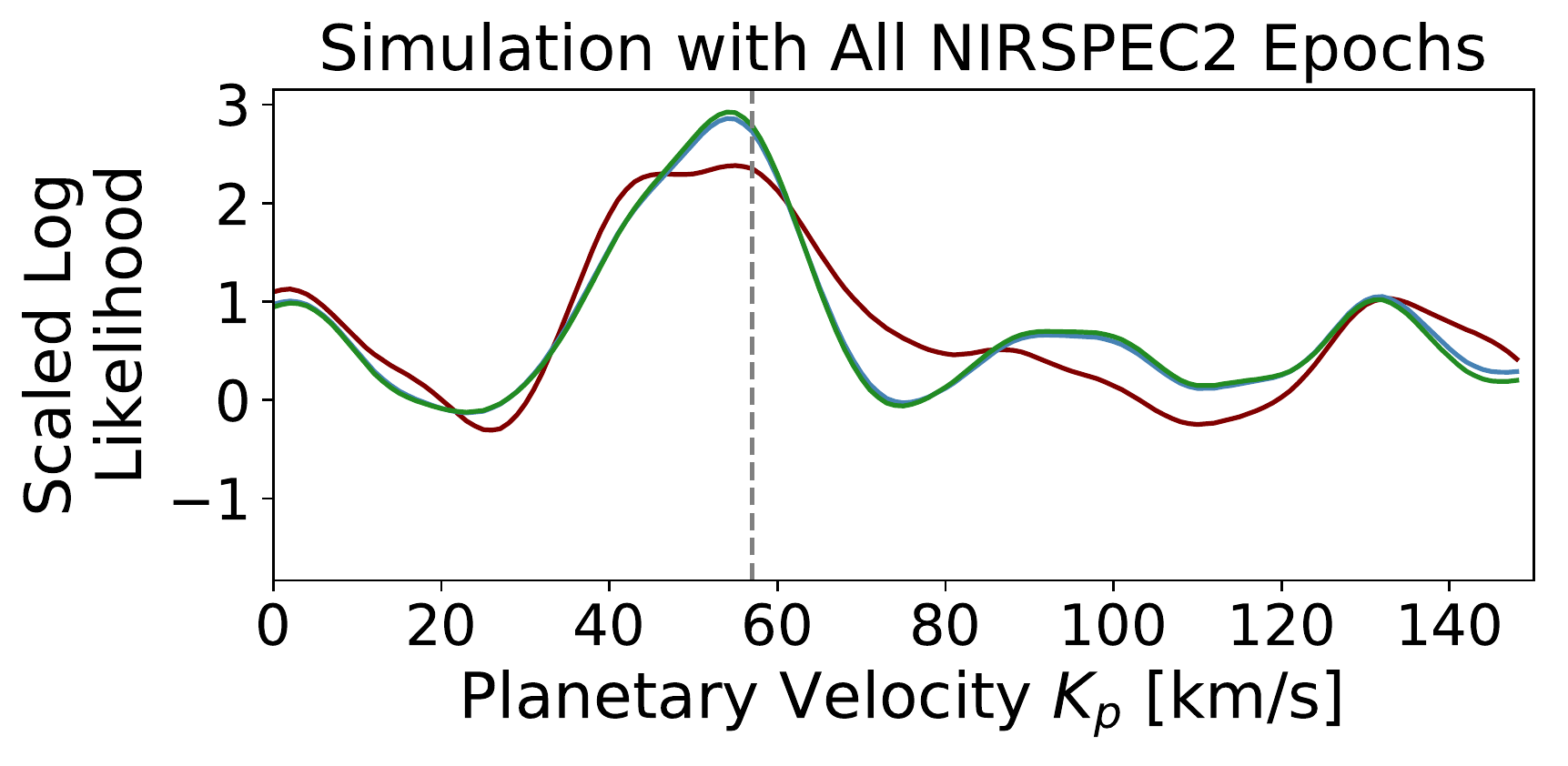}
    \caption{Similar to Figure~\ref{maxlike}C, but seven NIRSPEC2 epochs rather than five NIRSPEC1 and two NIRSPEC2 epochs. The three combination approaches give much more similar results on this homogeneous data set than on the heterogeneous data set shown in Figure~\ref{maxlike}C. This suggests that while all of the methods can detect the true peak in a homogeneous data set, the Zucker log(L) approach (blue) performs better on heterogeneous data sets, like our HD187123b one.  }
    \label{nirspec2only}
\end{figure}

This suggests that the Zucker log(L) approach is better suited for heterogeneous data sets than either the Brogi \& Line or the Zucker ML methods are. To test this hypothesis, we simulate the seven data epochs but as a homogeneous data set, i.e. with all NIRSPEC2 epochs rather than with five NIRSPEC1 and two NIRSPEC2 epochs. The NIRSPEC version determines the number of pixels per order, the number of orders, the instrument resolution, and the exact wavelength regions covered. We leave the S/N per epoch, planetary orbital phases, and barycentric velocities the same as in the real data set. Figure~\ref{nirspec2only} shows that with a homogeneous data set the two Zucker methods and the Brogi \& Line approach give much more equivalent results than they do with a heterogeneous data set, though the Brogi \& Line method still shows a side peak at $\sim$44 km/s that is not in the Zucker results. In other words, the Brogi \& Line approach is more sensitive to unwanted star/planet correlation than the Zucker approaches when applied to homogeneous data sets, but this effect is exaggerated with heterogeneous data. The Brogi \& Line log(L) function contains the variance of the data, which suggests that it should account for the variable noise across orders and epochs. Because of this, it may be surprising that it seems to perform worse on the heterogeneous data set than the Zucker log(L) method does. However, the make-up of each epoch (e.g. the specific wavelength range covered, the instrument profile, the orbital position, the barycentric velocity) could affect the level of per-epoch structured noise (e.g. planet/star correlation), to be reported on in future work. While the Brogi \& Line formalism accounts for differing levels of random noise between the epochs, it does not account for differing levels of structured, non-random noise. This could explain why it may not be performing as well on the heterogeneous data set as we may have expected it to. 

Because the Zucker log(L) method seems to produce the best results for our heterogeneous data set, we use it moving forward. We do, however, stress that further simulations of both different systems and inclination angles and heterogeneous data sets (different wavelength regions, different instruments) are needed to assess the robustness of log likelihood combination approaches.





\begin{figure}
    \centering
    \noindent\includegraphics[width=21pc]{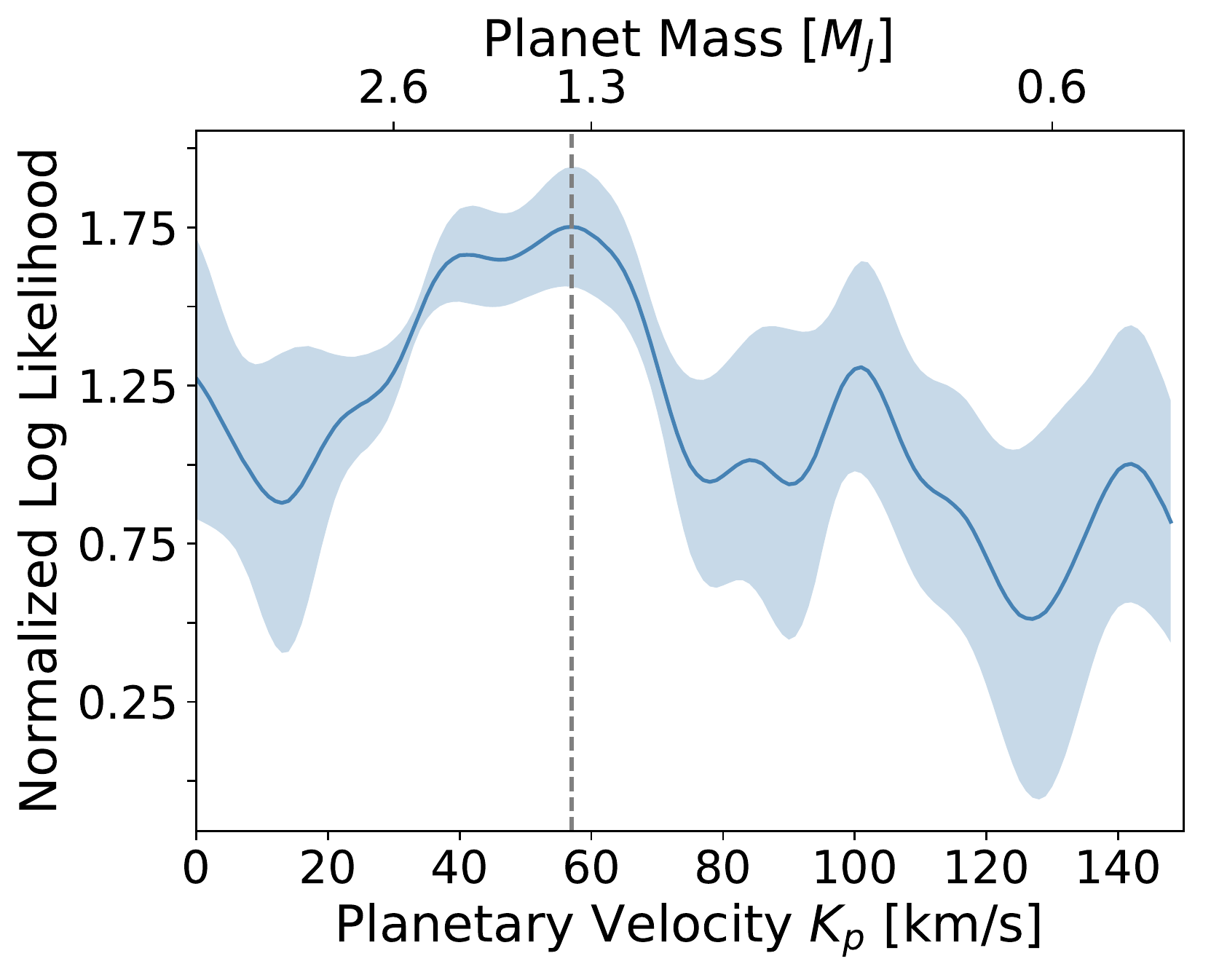}
\caption{Normalized log likelihood as a function of Keplerian orbital velocity $K_p$ for the HD187123b data using the Zucker 2003 log(L) CC-to-log(L) approach. The normalized log likelihoods plotted here and in subsequent figures are normalized by subtracting the mean of the log(L) from -150 to 150 km/s and adding 1. The curve shows the data results with the shaded region indicating the uncertainty ranges resulting from a jack-knife analysis of the data.    }
    \label{finalmaxlikeresults}
\end{figure}

To further investigate the validity of the peak at 57 km/s, we fit the simulations (as in panels of B and C of Figure~\ref{maxlike}) to the data and report the standard likelihood function
\begin{equation}
    \log L = \Sigma_k \bigg{(}\log\frac{1}{\sqrt{2\pi}\sigma_k}-\frac{(M_k-D_k)^2}{2\sigma_k^2}\bigg{)}
\end{equation}
where $M_k$ are the simulated pixels, $D_k$ are the data result pixels, and $\sigma_k$ are the uncertainty on the data results by pixel. To estimate our uncertainty on $K_p$, we use jack-knife sampling. Jack-knife sampling involves sequentially removing one epoch of data from the combination. The error is then equal to the $\sqrt{N-1}\ \times$ standard deviation of the N different combinations (where N is the total number of epochs). The jack-knife error bars are shown on the Zucker log(L) curve in Figure~\ref{finalmaxlikeresults}. As described in \citealt{piskorz88133}, jack-knife sampling is only one way of estimating error, which often actually overestimates the error because high variance between jackknife samples drives a high standard deviation, which produces large error intervals. Before fitting the simulations to the data, we normalize the simulated results by the ratio of the standard deviation of the data results and the standard deviation of the simulated results to account for the magnitude difference resulting from the uncertainty in planetary radius and lapse rate. A more sophisticated way of treating structured noise, for instance a Gaussian processes approach, is not yet computationally feasible for such high resolution data sets.
 
\begin{figure}
    \centering
    \noindent\includegraphics[width=21pc]{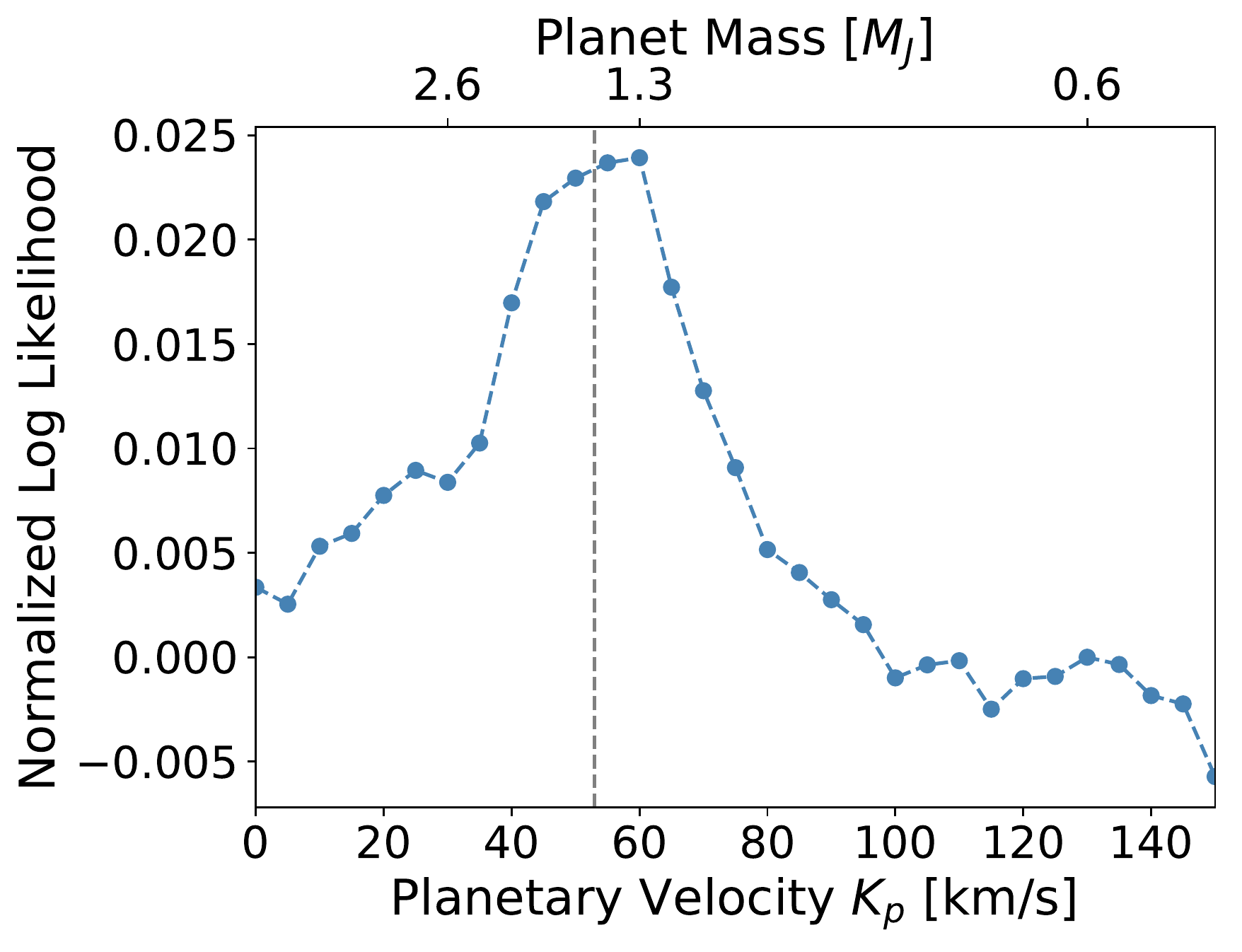}
\caption{Normalized log likelihood as a function of Keplerian orbital velocity $K_p$ between the data results and the simulated results using the Zucker log(L) cross correlation combination approach.     }
    \label{simulationfit}
\end{figure}

We test simulations from 0 to 150 km/s in steps of 5 km/s. The normalized likelihood is shown in Figure~\ref{simulationfit}. Fitting the simulations to the results allows us to remove unintended structure in the likelihood surface. In comparing the data result, shown in light blue in Figure~\ref{finalmaxlikeresults} to the likelihood result in Figure~\ref{simulationfit}, we can see how much of the unwanted structure, including that near 0 and between $\sim$90 and 150 km/s, is depleted. This indicates that the extraneous structure is not random, and can be removed by simulating multi-epoch data sets.  

To determine the uncertainty on $K_p$, we fit Gaussian functions to the results of each the raw data and the simulation fit to data results. From the raw data (shown in light blue in Figure~\ref{finalmaxlikeresults}), we measure a $K_p$ of 57 $\pm$ 15 km/s from the Zucker 2003 log(L) approach, while the simulation fit (Figure~\ref{simulationfit}) yields a $K_p$ of 53 $\pm$ 13 km/s. 

We determine the significance of the detection from the likelihood fit between the simulations and the data results, i.e. the function shown in Figure~\ref{simulationfit}, since real structure is minimized here and we can assume the variation at the baseline is from unstructured noise. We determine the noise level from the standard deviation of points beyond 2$\sigma$ from the peak. This gives a significance of 6.5$\sigma$ at 53 km/s. 

Previous multi-epoch detection works \citep[e.g.,][]{Piskorz2018} have reported significance by comparing the likelihood of a Gaussian fit (representing a detection) vs. a linear fit (representing a non-detection) to the peak. This method has given significances of hot Jupiter detections in the range of 3-4$\sigma$. This method was used previously because it was clear that the structure at off-peak velocities was not random and so an accurate noise level could not be obtained from it. Applying this technique to the raw data result, we measure a 3.6$\sigma$ detection from the Zucker log(L) approach. However, we were able to reduce the level of non-random off-peak structure, which allows us to determine the significance in a more straightforward way. While the two values of significance are not directly comparable, we do find a large increase in detection confidence by using simulations to correct out real off-peak structure.  




This $K_p$ of $53 \pm 13$ km/s corresponds to a planetary mass of $1.4^{+0.5}_{-0.3}\ M_J$ and an orbital inclination of 21 $\pm$ 5$^o$ at 6.5$\sigma$. We correlate the data with planetary models containing the spectral lines of only one molecule (H$_2$O, CO, or CH$_4$) and find that the log(L) surface is completely made from correlation with water lines. Therefore, we also report the 6.5$\sigma$ detection of water in the atmosphere of HD187123b. The log(L) curves produced from CO and CH$_4$ spectral models do not show peaks at the true $K_p$. This is not surprising, however, because CO does not have any spectral lines in the $L$ band wavelengths our data cover, and equilibrium chemistry predicts CO as the major carbon-bearing species in hot Jupiter atmospheres rather than CH$_4$.

\section{Signal to Noise vs. Orbital Coverage}
\label{snsection}


\subsection{Signal to Noise per Epoch}

\begin{figure}
    \centering
    \noindent\includegraphics[width=21pc]{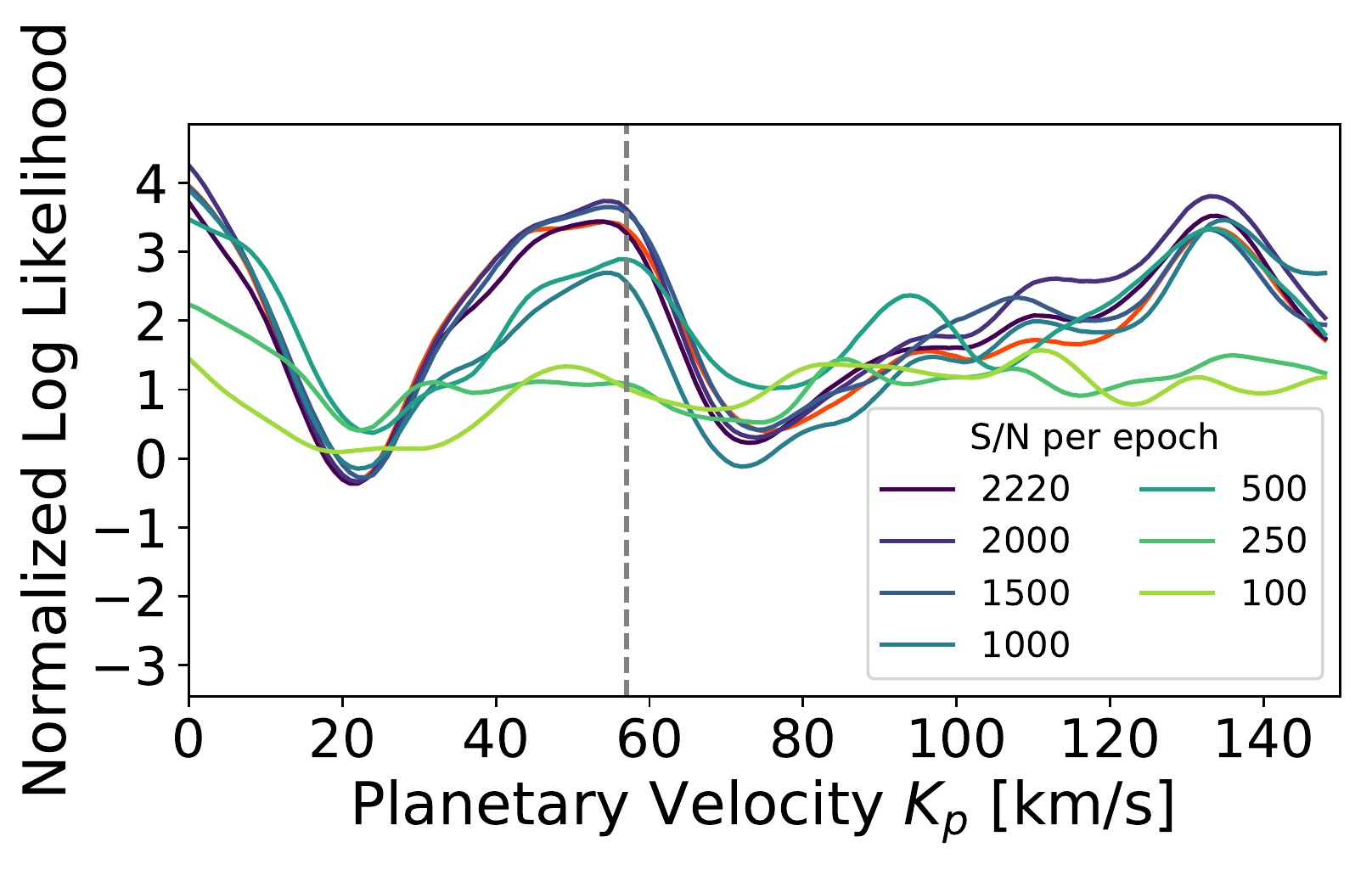}
\caption{Simulations showing how the $K_p$ detection decreases with decreasing S/N per epoch. The S/N is evenly distributed across the seven epochs. The 2220 S/N per epoch simulation has the same total S/N as the data results (shown in orange) that have an uneven S/N distribution, as described in Table~\ref{observationtable}. The similarity between the black, 2220 S/N per epoch curve and the orange curve demonstrated that the different distributions of the total S/N does not have a large effect on the structure of the final results.     }
    \label{snperepoch}
\end{figure}

The simulations used to fit the data (the results of which are shown in Figure~\ref{simulationfit}) elucidated the true planetary peak by reducing off-peak structure from correlation between the planetary and stellar spectral models. Though we could reduce this structured noise to a large extent, the detection significance is far from shot noise limited. Since this is the case, we investigate how the planet detectability would change with lower S/N epochs. To do so, we run simulations with the same parameters in the HD187123b data set described in Table~\ref{observationtable}, but decreasing S/N per epoch. To simplify these simulations we spread the total S/N evenly across the seven epochs, so each epoch has a S/N of 2220 to make up the total S/N of 5874 that we obtained in the data. The even distribution of S/N across epochs does not change the results much from the S/N distribution measured in the data as can be seen by the orange (data-like S/N distribution) and black (even S/N distribution) curves in Figure~\ref{snperepoch}. The rest of the curves in the figure show decreasing S/N per epoch. Interestingly, we see that the S/N per epoch can be degraded from 2220 per pixel to 1500 without any noticeable change in the height of the likelihood peak. Furthermore, the off-peak structure also remains the same until the S/N has degraded beyond a S/N of $\sim$500, confirming that this structure is real and not the result of random noise. 

\begin{figure}
    \centering
    \noindent\includegraphics[width=21pc]{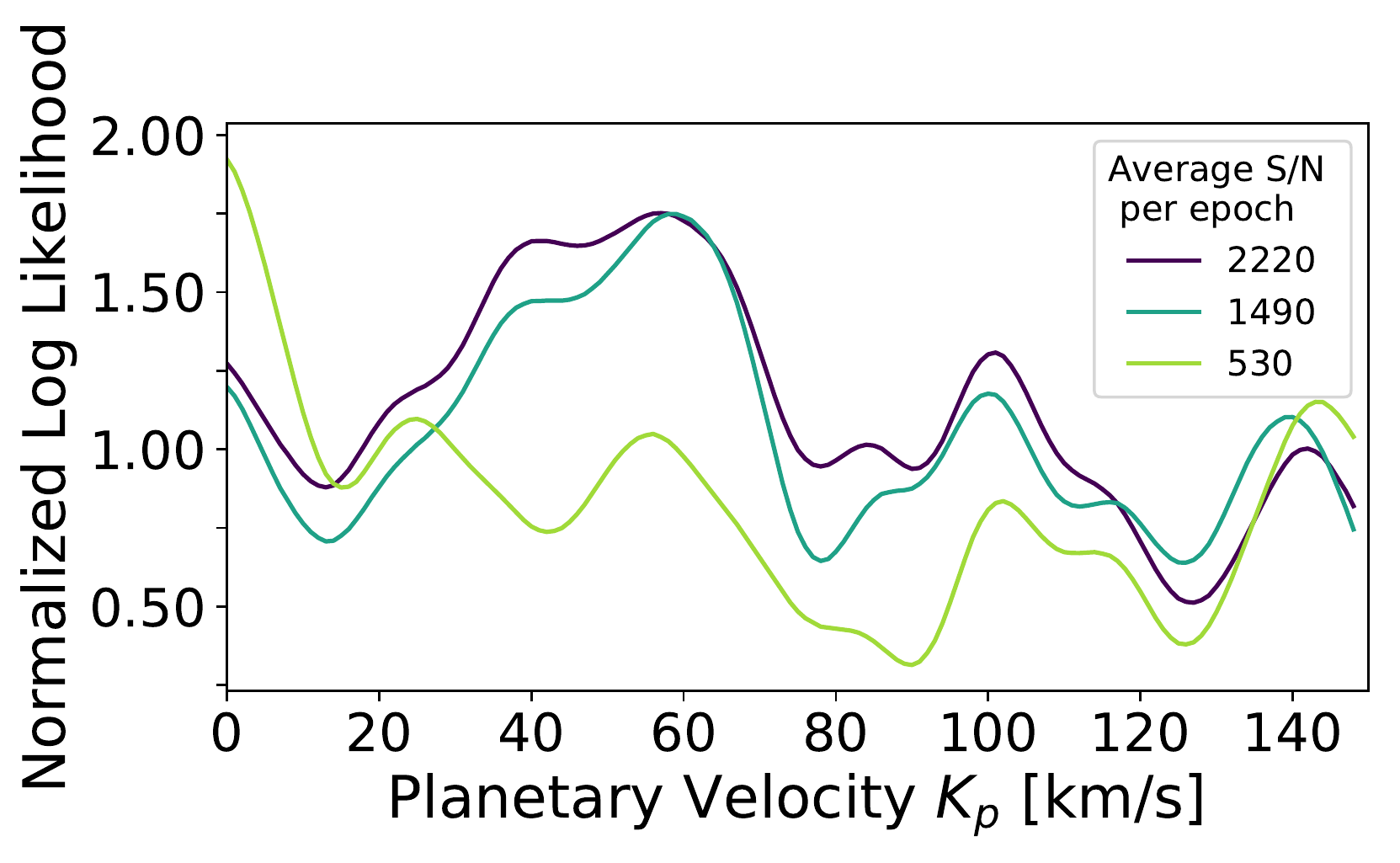}
\caption{Normalized log likelihood as a function of $K_p$ showing how chopping the data into lower S/N epochs affects the detection. The purple curve shows the results of the data with its full S/N per epoch, and the teal and green curves show the results when the data is chopped up such that there is an average S/N per epoch of 1490 and 530 respectively. As in the simulations shown in Figure~\ref{snperepoch}, we can see that the data epochs can be dropped from 2220 to $\sim$1500 while retaining quite similar peak and off-peak structure.       }
    \label{choppeddata}
\end{figure}

To further test these results, we chop the data into lower S/N epochs and test whether we see the same trend. By reducing the number of nods per epoch, we diminish the data set to seven epochs with average S/N per epoch of 1490 and 530 as well as the full average 2220 per epoch. We run PCA to remove telluric contamination after chopping the data, to approximate the results if we had truly only obtained the seven 1490 or 530 S/N epochs. In Figure~\ref{choppeddata}, the data set with 1490 S/N epochs gives produces a very similar shape to the full 2220 S/N epochs. The green curve, representing an average S/N of only 530 per epoch, also shows similar off-peak structure, for instance around $\sim$100 and $\sim$140 km/s, but the real peak is much diminished here. These results agree with those found using simulations, as seen in Figure~\ref{snperepoch}. These results, in both the simulations and the data, suggest that indeed, our detection is not shot-noise limited, and shorter epochs could be as effective for detecting planetary emission.

One feature seen in the data that is not seen in the simulations is the increase toward 0 km/s in the average S/N 530 epoch case. This set only considered two nods, which is the minimum possible to run a PCA-based telluric correction. Without a large offset in time between the two nods, there would not be as much change in the tellurics (airmass, abundances, plate scale, etc.), meaning that PCA could not remove the telluric contamination as effectively as it could in the higher S/N, more nod cases. The increase toward 0 km/s in the green curve is likely from correlation between the planetary spectral model and telluric contamination in the data. This sets a limit on how short the exposure time per epoch can be as long as a PCA-based approach is used to remove telluric contamination.

\subsection{Orbital Coverage}

We have seen that currently our detection confidence is limited by structured noise resulting from the correlation between stellar and planetary spectral models. Because we are not in the shot-noise limited case, and could achieve similar detections with lower S/N epochs, we test whether there is a more efficient way of using the full S/N that could help to remove the off-peak structured noise.

To test how we can reduce this structure, we run a simulation with the same total S/N as we obtained in the 7 data epochs, but instead we spread that S/N evenly across 20 epochs. These 20 epochs are evenly spaced across the orbit and with primary velocities evenly spaced between the maximum ($v_{sys} - $min($v_{bary}$)) and minimum ($v_{sys} - $max($v_{bary}$)). They have a S/N of 1313 per epoch as opposed to the average 2220 per epoch in the data. We use the NIRSPEC2 wavelength coverage and resolution to create the twenty epochs. 

\begin{figure}
    \centering
    \noindent\includegraphics[width=21pc]{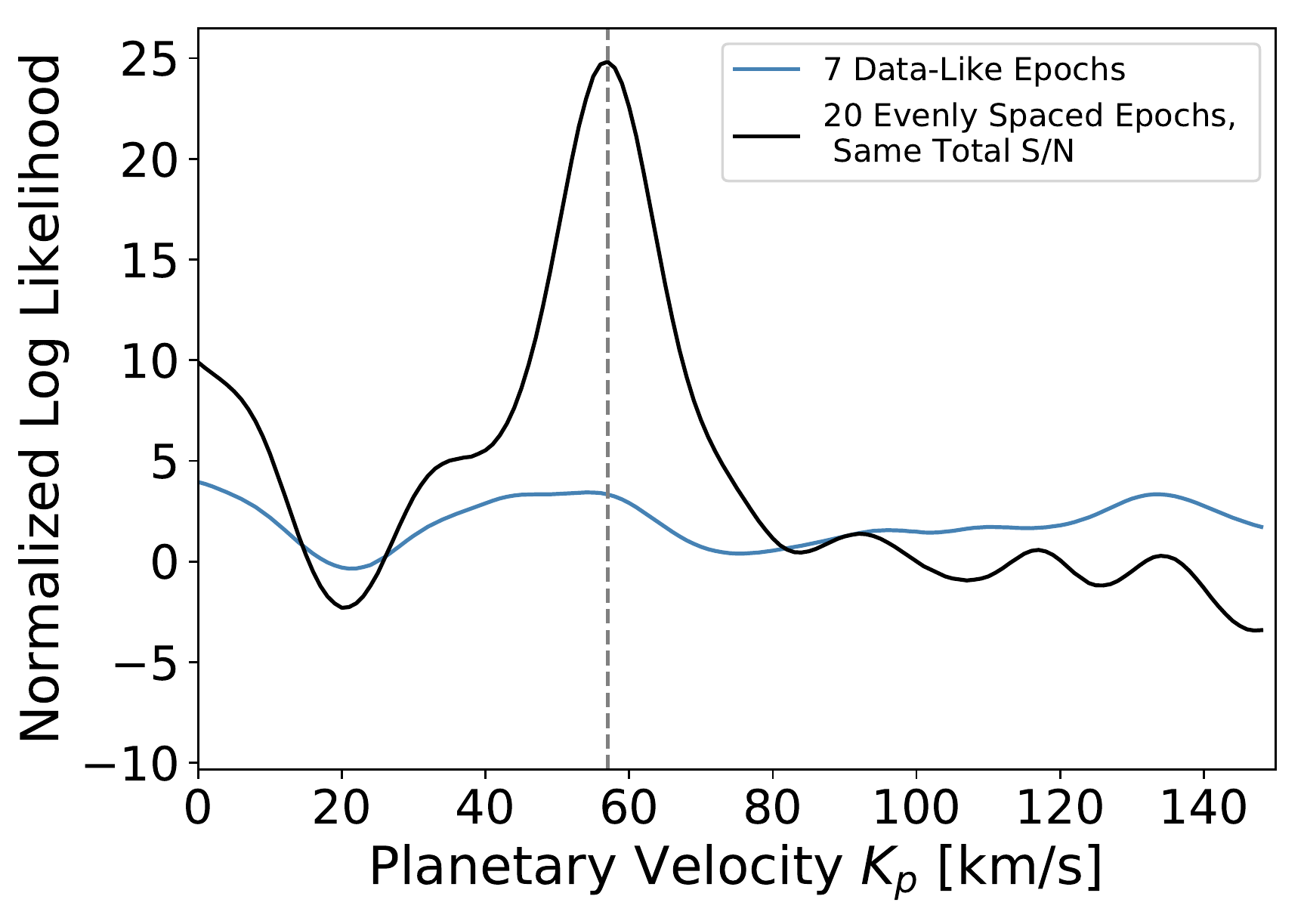}
\caption{Simulations showing the trade-off between S/N per epoch and number of epochs. The blue curve represents the simulations approximating our data set: with the same S/N per epoch, number of epochs, and epoch orbital positions. The black curve shows the results of simulations with 20 epochs evenly spaced across the orbit, but the same total S/N. The much stronger peak in the black curve implies that more, lower S/N epochs, i.e. greater orbital coverage, would give a much stronger detection than fewer, higher S/N epochs.         }
    \label{twentyepochs}
\end{figure}

The results of these simulations are shown in Figure~\ref{twentyepochs}. The blue curve represents the data-like simulations and the black curve shows the results of the twenty epoch simulations. Clearly, the twenty epochs result in a much stronger detection than do the seven epochs, even with the same total S/N. More epochs give us access to different wavelength shifts between both (1) the planet and the star and (2) the planet and the Earth's atmosphere, thus significantly reducing the correlation between the planet and star spectral models. It also reduces the the amount of the planetary spectrum that is lost to saturated tellurics because wavelengths that are lost to saturated tellurics will vary as the planet moves around its orbit and its spectrum is Doppler shifted relative to the stationary telluric lines. These simulations suggest that it would be more effective to spread the same total S/N over many epochs across the orbit rather than to obtain just a few isolated higher S/N epochs.

\section{Discussion}
\label{discuss}

The multi-epoch technique is a promising method for studying hot Jupiters and, in the future, cooler, further separated exoplanets, including those in habitable zones. It can access a much wider sample of planets that the transit technique can, and does not require the quickly changing line-of-sight planetary velocity that the single night technique does, or the spatial separation that  direct imaging programs do. Multi-epoch detections are currently limited by structured noise arising from correlation between the planetary models and the stellar component of the data. In this work, we investigate several ways of trying to reduce this unwanted structure.

The multi-epoch technique falls under the category of high resolution cross correlation techniques that must combine information from cross correlations of different segments of data. \citealt{zucker2003} and \citealt{brogi2019} each presented ways to convert cross correlations to log likelihoods so that they can be combined. We find that, for this heterogeneous data set, the \citealt{brogi2019} version gives more weight to the unwanted planet/star correlation at $\sim$44 km/s than either of the two \citealt{zucker2003} versions do. This suggests that the \citealt{zucker2003} combination method is better suited than the \citealt{brogi2019} for the two-dimensional cross correlation used in the multi-epoch technique, particularly for heterogeneous data sets (consisting of epochs with different resolutions, wavelength regions, number of orders, etc.). Future work comparing the three combination versions on two-dimensional cross correlations would be useful for really understanding the benefits and weaknesses of each technique, and for determining which would provide the strongest multi-epoch results moving forward.

We also present simulations that can reproduce the off-peak structure in the multi-epoch detection of HD187123b. We find that the detection is far from shot-noise limited and that in both simulations and data, the S/N per epoch could be reduced from 2220 to 1500 without a significant change in the shape of the normalized log likelihood vs. $K_p$ curve. We see that if we obtained many, lower S/N epochs rather than a few, higher S/N epochs, there would be a large increase in detection confidence, even without needing to fit the data results with simulated results.     

Being able to obtain useful information from lower S/N epochs could actually have a large impact on multi-epoch observing strategy. Since S/N increases with the square root of time, pushing from 2220 S/N epochs to 1500 S/N epochs, or from a total S/N of 5874 to 3968 per resolution element, we could save a factor of 2.2 in time. This suggests that a more traditional stellar radial velocity observation approach, such as a dedicated program on a smaller ground based telescope that could obtain many lower S/N epochs of data from many hot Jupiter systems, could be successful. 

The multi-epoch technique aims to learn about the bulk and atmospheric properties of exoplanets through directly detecting their Keplerian line-of-sight orbital velocity, $K_p$. More confident and constrained measurements of $K_p$, obtained through data sets with many, lower S/N epoch data sets, would provide more precise measurements of mass and inclination. Additionally, confident detections of $K_p$ will be critical for using multi-epoch detections to constrain atmospheric parameters, including metallicity and C/O. \citealt{oberg2011} found that, for giant planets that form via core accretion, the C/O ratio of the planet's atmosphere could be an indicator of whether it formed beyond the water snowline, where the gaseous C/O ratio is enriched relative to the stellar value, or within the water snowline, where the gaseous C/O ratio equals the stellar value. Such a measurement for a system like HD187123, with both a hot Jupiter and a Jupiter-analogue, could help to elucidate the processes of planetary formation and migration. 

We do note that a C/O measurement would likely require either $K$ or $M$ band data, in addition to the $L$ band data presented here, as the $L$ band contains H$_2$O lines while the $K$ and $M$ bands have prominent CO features. Future work to investigate whether many, lower S/N epochs could similarly improve $K$ and $M$ band detections, and how these improvements would affect constraints on C/O would be illuminating.




\section{Conclusion}
\label{conclude}
In this paper, we present a simulation framework that enables us to reduce the structured noise from multi-epoch direct detection campaigns (as in \citealt{lockwood,piskorz88133,piskorzupsand,Piskorz2018}) and elucidate the true planetary detection. Using this framework, we report the 6.5$\sigma$ detection of the thermal emission from the hot Jupiter HD187123b, and constrain its Keplerian orbital velocity to 53 $\pm$ 13 km/s. This allows us to measure the true planetary mass and orbital inclination of 1.4$^{+0.5}_{-0.3}\ M_{J}$ and 21 $\pm$ 5$^{\circ}$, respectively. We also report the presence of water in its atmosphere. We use these data sets to compare three methods of converting cross correlations to log likelihoods in order to combine them \citep{zucker2003,brogi2019} on multi-epoch data, and show that the Zucker log(L) approach is least affected by unwanted planet/star correlation for this data set. We also show that an observing strategy that spreads the total S/N across a planet's orbit rather than isolating it into a few, higher S/N epochs would inherently reduce this unwanted structure. The simulation framework presented here, and the optimized observing strategies it will permit, could provide a path from the atmospheres of non-transiting hot Jupiters down to those of habitable zone, Earth-sized planets.  



\acknowledgments{This manuscript benefited from conversations with Heather Knutson and from feedback from the referee, Matteo Brogi. The authors wish to recognize and acknowledge the very significant cultural role and reverence that the summit of Mauna Kea has always had within the indigenous Hawaiian community. We are most fortunate to have the opportunity to conduct observations from this mountain. The data presented herein were obtained at the W. M. Keck Observatory, which is operated as a scientific partnership among the California Institute of Technology, the University of California and the National Aeronautics and Space Administration. The Observatory was made possible by the generous financial support of the W. M. Keck Foundation. This work was partially supported by funding from the NASA Exoplanet Research Program (grant NNX16AI14G, G.A. Blake P.I.). }


\appendix
\section{Combining Cross Correlations} \label{crosscorrelations}

As high resolution cross correlation (CC) spectroscopy becomes more and more widely used to detect and characterize exoplanets, the questions of how to combine both (1) different segments of high resolution data and (2) high (e.g., NIRSPEC, CRIRES) and low (e.g., Spitzer, JWST) resolution data become important. \citealt{zucker2003} introduced an approach to convert cross correlations to log likelihoods (CC-to-log(L)) that can be applied in two ways. We will call these two versions of the \citealt{zucker2003} approach (1) the Zucker log(L) method and (2) the Zucker maximum likelihood or ML method. Previous multi-epoch detections of hot Jupiters \citep{lockwood,piskorz88133,piskorzupsand,Piskorz2018} have used the Zucker ML method. \citealt{brogi2019} recently presented a new CC-to-log(L) routine.

In this work, and for the multi-epoch technique in general, we use two-dimensional cross correlations (2D CC) to detect the unchanging stellar and planetary velocities during each epoch (see Section~\ref{corr}). Once the 2D cross correlations are calculated, we test each of the three different approaches to converting these cross correlations to log likelihoods. We first describe how the 2D CC is calculated, and then describe each of the approaches to converting these 2D cross correlations to log likelihoods.  

\subsection{One- and Two-Dimensional Cross Correlations}

When there is only one dominant spectral component in the data, the data can be described by the model
\begin{equation} \label{onedmodel}
    f(n) = ag(n-s)+d_n
\end{equation}
where $a$ is a scaling factor, $g(n)$ is a template spectrum in the same reference frame as the data, $s$ is a wavelength shift, and $d_n$ is the noise at bin $n$. In this case, a one-dimensional cross correlation function $C(s)$ is sufficient to match the model to the data and can be computed as
\begin{equation} \label{1dcc}
    C(s) = \frac{\Sigma_nf(n)g(n-s)}{N\sqrt{\sigma_f^2 \sigma_g^2}} ,
\end{equation}
where $f(n)$ and $g(n)$ are the target and template spectra, respectively, and the variances of the target ($\sigma_f$) and the template ($\sigma_g$) are given by
\begin{equation} \label{1Dsigma}
    \sigma_i^2 = \frac{1}{N}\Sigma_ni^2(n) .
\end{equation}
When there is more than one spectral component in the data, however, as is the case in the multi-epoch technique, the model described by Equation~\ref{onedmodel} can no longer accurately describe the data. Rather, a model considering two components is necessary,
\begin{equation}
    f(n) = a [g_1(n-s_1) + \alpha g_2(n-s_2)] + d_n
\end{equation}
As above, $a$ is a scaling factor and $d_n$ is the noise at bin $n$. The two spectral templates are given by $g_1$ and $g_2$ with wavelength shifts of $s_1$ and $s_2$, respectively. The scaling factor $\alpha$ accounts for the intensity ratio between the two template models. For this work, we set $\alpha$ equal to 0.0014, which is the spectroscopic contrast given by our stellar and planetary models and assuming a planetary radius of 1 $R_J$. We have found, however, that the {\em shape} of the resulting log likelihood surfaces, from both data and simulations, are independent of $\alpha$ in the range of $1.4 \times 10^{-3}$ to 10$^{-9}$. This is consistent with what was seen by \citealt{lockwood} and \citealt{piskorz88133}. 

\citealt{zuckertodcor} showed that a 2D CC $R(s_1,s_2,\alpha)$ could be calculated as  
\begin{equation} \label{2dcc}
    R(s_1,s_2,\alpha)=\frac{\Sigma_nf(n)[g_1(n-s_1)+\alpha g_2(n-s_2)]}{N\sigma_f\sigma_g(s_1,s_2)}
\end{equation}
where $\sigma_f$ is the same as described above, but $\sigma_g(s_1,s_2)$ can now be calculated as
\begin{equation} \label{2Dsigma}
    \sigma_g = \sqrt{\sigma_{g1}^2+2\alpha \sigma_{g1} \sigma_{g2}C_{12}(s_2-s_1)+\alpha^2\sigma_{g2}^2} .
\end{equation}
$C_{12}$ is the correlation between the two templates.

In all of the CC-to-log(L) approaches described below, we combine 2D CCs rather than 1D CCs. This involves replacing ($C(s)$) with ($R(s_1,s_2,\alpha)$) and using $\sigma_g$ calculated by Equation~\ref{2Dsigma} rather than by Equation~\ref{1Dsigma}. 

Once we have calculated the 2D log(L) surface for each epoch, we reduce to the one dimensional log likelihood functions (e.g., as seen in Figure~\ref{mlcombine}) by taking a cut along the maximum stellar velocity, which we check matches the expected stellar velocity from the combined systemic and barycentric velocities.

\subsection{\citet{zucker2003} log(L) Approach}
First, all correlations from a single night (segments from all orders after the saturated tellurics are removed) are combined using the approach from \citet{zucker2003}. This considers the observed spectrum $f(n)$ and a model $g(n)$ with a scaling factor ($a$), a shift ($s$), and random white Gaussian noise ($\sigma$). Expressions for $a$, $\sigma$, and $s$ can be found that maximize the log(L) between the observed spectrum and the model(s). By substituting these expressions in to log(L) equation, \citealt{zucker2003} showed that cross correlations can be related to log likelihoods (log(L)) as
\begin{equation}
    \log(L) = -\frac{N}{2}\log(1-R^2) \label{zuckerlogLequation}
\end{equation}

The individual cross correlations are converted to log likelihoods and summed for each epoch. The fact that the cross correlation $R$ is squared in this operation means that a negative correlation would provide the same log likelihood as a positive correlation. In other words, a model would give the same log likelihood when fit to the data at a given velocity whether it were multiplied by 1 or -1. This is concerning because, while absorption and emission lines are not merely related by a sign-flip, correlation between an absorption line in the data with an emission line in the model, or vice versa, would produce an anticorrelation, which would be given the same likelihood as a corresponding positive correlation by Equation~\ref{zuckerlogLequation}. The pressure/temperature profile of a planet's atmosphere, whether inverted or non-inverted, determines whether lines will show up in absorption versus emission, and so not being able to distinguish between the two cases severely limits our ability to understand atmospheres. To account for this, we correct any negative correlation values to zero. This is done by calculating $\log(L)$ as
 \begin{equation}
\begin{array}{r}
y_i(s) =  \left\{
        \begin{array}{ll}
            N_i \log (1-R_i(s)^2) & \quad R_i(s) \geq 0 \\
            N_i \log (1+R_i(s)^2) & \quad R_i(s) < 0
        \end{array}
    \right\} \\
\log(L(s))= \left\{
        \begin{array}{ll}
            -\frac{1}{2}\Sigma_i  y_i(s)  & \quad \Sigma_i y_i(s) < 0 \\
            0 & \quad \Sigma_i y_i(s) \geq 0
        \end{array}
    \right\}
\end{array}     
 \end{equation}
Applying this correction after summing the $y_i$'s, rather than for each negative $R_i$, accounts only for heavily weighted negative correlations. That is, we do not set negative values in the individual $R_i$'s equal to zero before combining them because we wish to retain the information from negative $R_i$'s that arise from noise or uncertainty in the spectra. By waiting until the $y_i$'s are combined to make this cut, we avoid automatically losing both small negative values in the $R_i$'s or negative values in a $R_i$ that have very small relative weighting ($N_i$). This correction creates the horizontal portions at zero of the stellar log likelihood curve in Panel A of Figure~\ref{mlcombine}. This method of correcting negative correlations has been used in previous multi-epoch analyses \citep[e.g.,][]{piskorz88133,piskorzupsand,Piskorz2018}, and we describe it here for transparency.

Stellar lines are the dominant component of our data set and the real planetary signal must correspond with the correct stellar velocity. In other words, we can only detect the planetary signal once the model and data stellar lines are matched up. Therefore, the variation in the planetary correlation is around the mean stellar correlation peak, which is well above zero. As a result, and because we know that the stellar lines are in absorption rather than emission, negative correlation values only appear, and are corrected, at incorrect stellar velocities. Because the planetary correlation values will never reach down to, or below, zero, correcting negative correlation will not affect the planetary curves. Anticorrelation between the planetary lines in the data and model will be distinguishable from correlation between it will result in smaller (i.e. below the stellar correlation baseline), but still positive, correlation values.

We want to stress that negative correlations should not be corrected when using a one-dimensional cross correlation or when the two spectral components in a two-dimensional cross correlation are of similar strength. Doing so would artificially alter the distribution of likelihood values which would invalidate the uncertainties given by the resulting likelihood surface.

 
Then, the log(L) from different nights of data are converted from $v_{sec}$ to $K_p$ space according to Equation~\ref{vsec}. Finally, the log likelihoods are summed to find the most likely $K_p$. 

\subsection{\citet{zucker2003} ML Approach }

The Zucker ML method follows the Zucker log(L) method up to Equation~\ref{zuckerlogLequation}. However, rather than combining the likelihoods at this point, \citealt{zucker2003} show that individual correlations can be combined into an `effective' correlation value, ML, as follows:
\begin{equation}
    N_{tot}\log [1-\textrm{ML}^2(s)] = \Sigma_i N_i \log[1-R_i^2(s)]
\end{equation}
where the right side is the sum of the log(L)'s of individual segments and the left side is the log(L) of the full data set (from a single night where the planetary velocity is constant). The $R_i$'s and $N_i$'s are the 2D cross correlations and number of pixels of each of the segments, respectively, and $N_{tot}$ is the total number of pixels. By analogy, ML is the effective correlation of the full data set. Because ML is an effective correlation, we rename it $R(s)$ and evaluate it as,
\begin{equation}
    R(s) =  \sqrt{1-\exp\bigg{(}\frac{1}{N_{tot}}\Sigma_i N_i \log[1-R_i^2(s)]\bigg{)}}.
\end{equation}
This gives us an effective correlation for each epoch. We correct for negative correlation values here in an analogous fashion to that described for the Zucker log(L) approach. The effective cross correlations can then be converted to log(L) following \citealt{lockwood}:
\begin{equation}
    \log(L) = \mathrm{const} + R(s).
\end{equation}
Finally, the log(L)'s from different nights are converted from $v_{sec}$ to $K_p$ space, as in the other approaches, and summed. 

This was the CC-to-log(L) approach used in the previous NIRSPEC multi-epoch detection papers \citep{lockwood,piskorz88133,piskorzupsand,Piskorz2018}.

\subsection{\citet{brogi2019} Approach}
\citet{brogi2019} recently presented a new approach to converting cross correlations to log(L). Instead of substituting the expression for $a$ that maximizes the log(L) between an observed spectrum and a model, they set $a$ equal to 1. Setting $a$ to 1 allows for discrimination between correlation and anticorrelation, or between emission and absorption lines. We note that in our 2D case, where there are both stellar and planetary signals in the data, a negative $a$ would invert the stellar absorption lines as well as the planetary lines. Our data have high enough S/N on the stellar lines that flipping the stellar model would produce a strong anticorrelation, which would be corrected to zero as described above. Therefore, our data would not select for negative $a$ values and even without setting $a$ to 1, the Zucker methods would not run into the issue of confusing planetary (and stellar) emission and absorption lines.
 

By setting $a=1$, \citet{brogi2019} derive the expression
\begin{equation} \label{brogilogL}
    \log(L)=-\frac{N}{2}\bigg{\{}\log(\sigma_f \sigma_g)+\log\bigg{[}\frac{\sigma_f}{\sigma_g}+\frac{\sigma_g}{\sigma_f}-2R(s)\bigg{]}\bigg{\}}
\end{equation}


We stress that since our approach uses two dimensional cross correlations, $R(s)$ and $\sigma_g$ are the two dimensional variants described in Equations~\ref{2dcc} and \ref{2Dsigma}, rather than the one dimensional $C(s)$ and $\sigma_g$ described in Equations~\ref{1dcc} and \ref{1Dsigma}.  
 
As in the \citealt{zucker2003} approach, the log(L) functions from a single night are summed, then the summed log(L) for each night is converted from $v_{sec}$ to $K_p$ space and summed.

{\footnotesize
\bibliography{sample}}
\bibliographystyle{ApJ}

\end{document}